\documentclass[floatfix,nofootinbib,aps,prb,preprint,groupedaddress,showpacs,amssymb,amsfonts]{revtex4}
\usepackage{graphicx,epsfig,pstcol,pstricks,amssymb}
\usepackage{hyperref}
\hypersetup{colorlinks=true,urlcolor=blue}
\begin{document}
{\small Molecular Physics, volume {\bf 101} pp. 2241-2255 (2003)\\ \htmladdnormallink{http://dx.doi.org/10.1080/0026897031000112424}{http://dx.doi.org/10.1080/0026897031000112424}}

\setcounter{page}{1}

\title{THE PROPERTIES OF FULLY FLEXIBLE LENNARD-JONES CHAINS 
IN THE SOLID PHASE: WERTHEIM TPT1 THEORY AND SIMULATION}

\author{ Eduardo Sanz, Carl McBride and Carlos Vega }
\address{ Departamento  de Qu\'{\i}mica F\'{\i}sica,
   Facultad de Ciencias 
Qu\'{\i}micas, Universidad Complutense de Madrid.
 Ciudad Universitaria 28040 Madrid, Spain.  }

\begin{abstract}
NpT ensemble Monte Carlo simulations  were performed for
fully flexible Lennard-Jones chains in the solid phase. The bond length
between monomers within the chains is fixed to $L=\sigma$ and the 
molecule is free to adopt any configuration. 
The solid structure of fully flexible chains is obtained by
randomly locating  the bonds of the chain within a 
face centered cubic close packed arrangement of atoms.
For fully flexible chains it is believed that the stable solid
phase is  disordered. Such a solid is  considered in this work.
Computer simulations were performed for chains with $m=3, 4$ and 5 monomer
units, and results were obtained for the equation of state and
internal energy of the chains.
An extension of Wertheim's TPT1 to the
solid phase of Lennard-Jones chains (C.Vega, F.J.Blas and A.Galindo,
J.Chem.Phys.,{\bf 116},7645,2002) has been proposed 
recently. The simulation results of this work provide  a
check on the performance of this theory. It is found that 
Wertheim's TPT1 successfully predicts the equation of state and
internal energies of fully flexible LJ chains in the solid phase.
Finally, a rigid  LJ chain in a linear configuration has been considered.
Computer simulations were also performed for the rigid chain in 
an ordered solid structure. 
It is found that fully flexible
and linear rigid chains present  quite different equations of state 
and different thermodynamic
properties in the solid phase.
\end{abstract}
\maketitle
\pagebreak

  \section{INTRODUCTION}

 One of the simplest molecular models that one can conceive of for
flexible chains is that formed by tangent Lennard-Jones monomers.
The pair potential between two monomers (either belonging to the same
molecule or belonging to different molecules) 
is given by the Lennard-Jones
potential with potential 
parameters $\epsilon$ and 
$\sigma$. The reduced  bond length $L$ between bonded monomers
is fixed to the value $L^{*}=L/\sigma$ so that, in some respects,  one may say that
the chain is formed by "tangent" monomers.
There is neither a bending potential constraining the angle between
three contiguous monomers of the chain, nor a torsional potential
between four contiguous monomers. 
We shall take the liberty of denoting this model as the 
fully flexible tangent LJ chain model, or, more succinctly, as the LJ chain model.
The LJ chain model provides an ideal starting point for the computational and 
theoretical study of simple polymer chains.
One of the attractive features of the LJ chain model in such a study is the large and ever
growing body of theoretical and simulation work published over the last decade or so.
 
A great in-road to the theoretical treatment of molecular fluids was provided by Wertheim.
In the nineteen eighties Wertheim developed a theory for associating
fluids \cite{wer1,wer2,wer3,wer4}. 
  When the association strength becomes infinitely
 strong then chains are formed from a fluid of associating
 monomers \cite{wer5,chap1}. In this way 
it was possible to derive an equation
 of state for a chain of freely jointed tangent monomers 
 by using thermodynamic information of the monomer reference
 fluid only.  In the simplest implementation of the theory (TPT1)
 the only information required to build an approximate
 equation of state for the chain fluid is the equation of
 state and the pair correlation function at contact
 of the monomer fluid. Although Wertheim's formalism was originally
 conceived for hard chains it soon realized that it could also
 be applied to LJ 
chains \cite{chap2,johnson1,johnson2,blas1,blas2,binder,mullergubbins}.
 It has been shown that Wertheim's TPT1 provides a good
 description of the thermodynamic properties of the LJ chains
in the fluid phase (including the description of the vapor-liquid
 equilibria). 

 Although a lot of work has been devoted to the study of LJ
chains in the fluid phase\cite{sim1,sim2}, studies of 
LJ chains in the solid phase
are rather scarce. The solid phase of diatomic LJ models 
has been considered\cite{lisal} by 
Lisal et al. (for $L^{*}=L/\sigma=0.67$)
and 
by Vega, Blas, Galindo \cite{vega2002} (for $L^{*}=1$). Semi-flexible
LJ chains (with $m=6$ and a bending potential favoring the linear
configuration) have been considered by Polson and 
Frenkel\cite{polson1}.  A natural question that arises is the possibility
of extending Wertheim's TPT1 to the solid phase.
Recently we have shown that such an extension is indeed possible.
Vega and MacDowell \cite{vegamacdowell2001} have shown
that Wertheim's TPT1 can be extended to the
solid phase of freely jointed hard
spheres obtaining excellent agreement with the simulation
results of Malanoski and Monson\cite{malanoskichains}.
Similar results were obtained in two dimensions \cite{carl}.
Encouraged by these results Vega, Blas and Galindo extended 
Wertheim's TPT1 to LJ chains
in the solid phase\cite{vega2002}. It has been shown 
that Wertheim's TPT1 provides
a good description of the equation of state and internal 
energy of the LJ dimer in the solid phase. 

Theoretical predictions 
for longer LJ chains in the solid phase (formed by m=3,4,5,6 
monomers) have been published  \cite{vega2002}
but, due to the lack of simulation results for LJ chains in
the solid phase, a direct comparison with the numerical results
was not possible. The goal of this paper is to perform computer
simulations of LJ chains in the solid phase and to compare
the theoretical predictions of Wertheim's TPT1 to the numerical
results.

 Let us briefly discuss the solid structure of LJ chains. 
Lennard-Jones monomers freeze \cite{hansensolid,kofke} into a close packed 
structure (face centered cubic, fcc). In principle one may suspect
that the solid structure of LJ chains should be related to the
close packed structure of the LJ monomer. 
One possibility for the solid structure of LJ chains is to
form layers of molecules with the chains adopting a linear
configuration, with all the molecules within the same
layer pointing in the same direction. 
Examples of such  of structures were presented in earlier work
for hard diatomic 
fluids (hard dumbbells) by Vega et al. \cite{vega1992a}
and for LJ chains (with bending potential favoring linear
configurations) by Polson and Frenkel\cite{polson1}. 
In the particular case of the so called CP1 structure (see 
Ref.\cite{vega1992a} for details) the molecular
axis of all the molecules of the solid point in the same direction.
There is  no doubt that this should  indeed be the stable solid
structure when a "bending" potential between contiguous monomers
of the chain exists, as is the case in the study by Polson
and Frenkel, or for a linear rigid chain.

 However, it is likely that ordered structures (those formed
by layers of oriented molecules) are not the stable solid phase
for fully flexible LJ chains (i.e. chains with no bending potential).
It is possible to build a solid
structure where the atoms follow an face centered cubic (fcc)
close packing but
where the bonds of the dimers are located randomly within
the solid, with no long range orientational
order between the bond vectors of the chains.
Wojciechowski et al.\cite{branka2,branka1} were the first
to realize this important feature . We shall denote 
this structure as the disordered 
solid.  In fact Wojciechowski, Branka
and Frenkel \cite{branka2,branka1} showed that the
stable solid structure of tangent
hard discs dimers in two dimensions is formed by a close
packed arrangement of atoms with a disordered arrangement of
bonds. The same idea holds for hard chains in three dimensions
\cite{malanoskichains} and one may expect that the same
would occur for a three dimensional LJ fully flexible chain.

In a sense this is one of natures clever solutions. 
The molecules may achieve the
close packing structure of the monomers, leading to an optimally high density.
At the same time the bond structure of the system is disordered.
The great beauty of such a system is the stabilizing effect 
of an additional contribution
to the entropy of the system arising from the degeneracy of 
the solid structure (i.e. there are a number of ways of forming
the disordered solid whereas there is only one way of forming the
ordered solid). For m=2 this degeneracy has been estimated
by Nagle\cite{nagle}, and for longer chains Malanoski and Monson have shown
that can be estimated quite accurately from the Flory-Huggins
lattice theory of chains\cite{huggins1,huggins2}.

  The scheme of the paper is as follows. In Section II the
extension of Wertheim's theory to the solid phase of LJ
chains shall be  be described. In Section III we shall provide
details of the simulations performed in this work.
In Section IV the results of this work will be presented, and
in section V the conclusions shall be presented.

\pagebreak

 \section{BRIEF DESCRIPTION OF WERTHEIM'S PERTURBATION THEORY}

Since details of the extension of Wertheim's TPT1 to the solid
phase for  LJ chains has been published previously \cite{xxx} the following forms  only an outline description of 
the technique.

  Let one  assume that there are  a certain 
number, $N^{ref}$, of spherical monomer
 particles within a certain volume $V$ at temperature $T$.
 These spherical particles
 interact  with each other via  a spherical pair potential $u^{ref}(r)$. In this
 work the pair potential $u^{ref}(r)$ will be the Lennard-Jones 
 potential with
 parameters $\sigma$ and $\epsilon$. We shall
 denote this fluid as the reference fluid and the properties of
 this reference fluid will be labeled by the superscript $ref$.
 Let us also assume that
 in another container of volume $V$ and temperature $T$, we have
 $N=N^{ref}/m$  fully flexible chains of $m$ monomers each.
 By fully flexible chains we mean chains of $m$ monomers, with
 a fixed bond length of $L=\sigma$, and no other constraints (i.e there
 is no restriction in either bonding angles or in the torsional
 angles).  Each monomer of a certain chain
 interacts with all the other monomers in the system (i.e in the same
 molecule or  in other molecules with the only exception being the
 monomer/s to which it is bonded)
 with the pair potential $u^{ref}(r)$.
 The chain system described so far will be denoted as the chain fluid.
 It follows from Wertheim's TPT1 that the free energy of the
 chain model is given by:

\begin{equation}
  \frac{A}{NkT} = ln(\rho) -1 + m \frac{A^{ref}_{residual}}{N^{ref}kT}
-(m-1) \ln  y^{ref}(\sigma)
 \label{achain}
 \end{equation}

  The above equation shows that the free energy of the chain  may
  be obtained from a knowledge of
the residual free energy of the reference
  fluid and the pair background correlation function $y(\sigma)$ 
of the reference fluid at the bonding distance of the chain\cite{hansen,macquarrie}.
It is worth  recalling that Wertheim's TPT1 is normally used to 
to describe
chains formed by "tangent" spheres (i.e those with 
reduced bond length between monomers $L^{*}=1$). 
Since for the LJ potential model the pair potential is zero for
$r=\sigma$ it turns out that for this particular choice of 
bond length, $y(\sigma)$ is equal to $g(\sigma)$, so that
one is able to  replace $y(\sigma)$ by $g(\sigma)$ in Eq.\ref{achain}.

  The equation of state which follows from Eq.\ref{achain} is given by:

\begin{equation}
  Z=  m Z^{ref} - (m-1) \left( 1 + \rho^{ref}
\frac{ \partial \ln g^{ref}(\sigma) } {\partial \rho^{ref}} \right)
\label{zchain}
\end{equation}
\noindent where we have 
defined $Z^{ref}$ as $Z^{ref}=p^{ref}/(\rho^{ref}kT)$.
  The residual part of the internal energy U is given by :

\begin{equation}
  \frac{U} {NkT} = m \frac{U^{ref}} {N^{ref}kT} +
 (m-1) T \frac{ \partial \ln g^{ref}(\sigma) } {\partial T } 
\label{uchain}
\end{equation}

  We shall denote Eq.\ref{achain} , Eq.\ref{zchain}  and
Eq.\ref{uchain} as Wertheim's
TPT1 theory.
We note that the arguments used to arrive at Eq.\ref{achain} and
Eq.\ref{zchain} make no special mention whatsoever to the actual nature
of the phase considered \cite{sear,vegamacdowell2001,praus2002}. Therefore 
the equations can be applied  to both  the fluid and  the solid
phases.  The possibility 
of extending Wertheim's TPT1 
to the solid phase has been explored only in the last three
years \cite{vegamacdowell2001,carl,vega2002}. 
 All that is 
required in order  to obtain a fully unified
theory for the phase equilibria of chain
molecules is the residual free energy, compressibility factor and pair
correlation function of the monomer fluid both for the fluid and solid
phases. For the fluid phase Johnson et al.\cite{johnson1,johnson2} have 
provided values
of the free energy and structural properties (i.e $g^{ref}(\sigma)$) of
the monomer LJ fluid.
Therefore, for the fluid phase we shall use the implementation of
TPT1 provided by Johnson et al. \cite{johnson1,johnson2}.
  For the solid phase van der
Hoef \cite{hoef} has recently proposed an analytical
expression for the free energy of the LJ monomer solid.
This analytical expression is essentially a fit to the most recent
simulation results for the solid phase of this model.
Therefore we shall simply adopt the expression provided by van
der Hoef for the free energy of the LJ monomer solid. The other piece of
information required by the theory is the value of $g^{ref}(\sigma)$ for
the LJ monomer solid. In Ref.\cite{vega2002} we have 
performed computer simulations of the LJ
monomer in the solid phase 
to obtain $g^{ref}(\sigma)$ for a number of temperatures and
densities. The simulations results for $g^{ref}(\sigma)$ were fitted
to an empirical expression similar to that proposed by Johnson
et al. for the fluid phase \cite{johnson2}. In this work we 
shall use 
the expression
given by us in Ref.\cite{vega2002} 
for $g^{ref}(\sigma)$ of the solid phase, 
which is essentially a fit to the 
structural results from simulations of the LJ monomer solid.

\pagebreak

 \section{SIMULATION DETAILS}
The model considered in this work consists of a chain formed
by $m$ identical LJ sites. The bond distance between contiguous
monomer units is given by $L=\sigma$. 
The pair interaction between a pair of molecules
is given by :

\begin{equation}
 u(1,2) = \sum_{i=1}^{i=m} \sum_{j=1}^{j=m} 4 \epsilon 
\left( \left(\frac{\sigma}{r_{ij}} \right)^{12} - \left(\frac{\sigma}{r_{ij}}\right)^{6} \right)
\end{equation}
where $r_{ij}$ is the distance between site $i$ of molecule 1
and site $j$ of molecule 2.
We also used the LJ potential to describe the interaction between
monomers of the same chain separated by at least two molecular 
bonds
(i.e. a LJ potential was not included between contiguous bonded
monomers of the same chain).

In order
to describe a disordered structure, we generated a close packed
arrangement (fcc) of atoms with the molecular bonds randomly
distributed. A trial and error algorithm was used to generate the initial 
disordered structures. The algorithm starts at one corner of a face centered cubic lattice with 
sufficient number of lattice points to contain an integer number of molecules. Bonds
are made to unoccupied randomly chosen adjacent lattice points until a molecule is placed. The algorithm then 'looks' 
for the next vacant space in order to place the first atom of the second molecule. Once placed,
bonds are once again made by randomly selecting  one of the unoccupied adjacent lattice points. 
If none of the adjacent lattice points are unoccupied then the algorithm restarts from the beginning.
This algorithm only works well for short chains with a modest number of lattice points.
Once a randomly ordered configuration has been generated then the structure can be 
copied, rotated  and translated. In this work the initial structures were as follows:
for $m=3$, 64 molecules were randomly packed, for $m=4$, 27 molecules were
packed  randomly, and for $m=5a,$ 18 molecules were randomly placed.
These initial structures were then duplicated twelve times for $m=3$
thus resulting in a system with $N_{m}=2304$, eight times for $m=4$
thus resulting in a system with $N_{m}=864$, and eight times for $m=5$
thus resulting in a system with $N_{m}=720$.

For a number of representative systems a second disordered configuration was generated
in order to check for differences in the thermodynamic properties. Upon analysis such
differences were found to be minimal.
Simulations were initiated at very high pressures where the
density is close to the close packing limit (note that there is no
true close packing density for a soft potential such as the LJ case, however,
the reduced monomer number density of hard spheres at close packing 
$\sqrt{2}$ proved to be a good starting point). After generating
the initial structure at the close packing density (i.e.
the reduced number density of LJ monomers was $\sqrt{2}$) 
this structure was expanded to lower densities by performing
NpT Monte Carlo simulations \cite{allen} at successively lower pressures.
Since the distribution of bonds in the solid phase was 
isotropic, changes in the volume of the simulation box were made isotropically.

For  all of the simulations performed in this
work the site-site LJ pair potential was truncated 
at $r_{c}=2.5\sigma$ and long range interactions were
added to all the computed thermodynamic properties (internal energy, pressure) 
by assuming
that the site-site pair correlation function is unity beyond the
cutoff\cite{allen}. A cycle will be defined as a trial move per particle,
plus a trial volume change. Three different kind of trial
moves were used: translation of the whole molecule, rotation of 
the whole
molecule and configurational bias move. The aforementioned trial moves were performed with the 
following probabilities; 
$40\%$,$40\%$ and $20\%$ respectively. In the configurational
bias move a segment of the chain was chosen randomly and
the chains were re-grown in a random direction. Typically 
$N_{k}=8$ trial orientations were used for re-growing each segment of the
chain. Since configurational bias is now a standard technique
we refer the reader to \cite{frenkelsmit} for further details. 
Throughout  this work 
reduced units shall be used,
so that $T^{*}=T/(\epsilon/k)$, 
$\rho^{*}=\rho \sigma^{3}= (N/V) \sigma^{3}$, 
and $p^{*}=p/(\epsilon/\sigma^{3})$.  
A typical run 
of the solid phase involved 
70000 cycles of equilibration followed by 70000 cycles for
obtaining equilibrium properties. 

In summary,  isotropic NpT ensemble was applied to an  fcc arrangement of
atoms with randomly  assigned  bonds vectors, using translation
rotation and configurational bias moves. 

\section{RESULTS}

As an initial check of the simulation code and the structure generation algorithm 
a system of fully flexible hard sphere chains with $m=4$ monomers was simulated.
The results of this simulation were compared to the work of 
Malanoski and Monson\cite{malanoskichains}.  The results of this benchmark 
are presented in Fig.1. 
As  can be seen the simulation results of this work agree
rather nicely with those of Malanoski and Monson. Along with the simulation results
 the predictions
from Wertheim's TPT1 for hard sphere chains \cite{vegamacdowell2001}
in the solid phase
are also presented. As it can be seen the agreement is again quite good.
We shall not comment further  on this point since extensive comparison
between simulation and theory for hard sphere chains  exists in published literature \cite{xxx}.

 In tables I,II and III simulation results for LJ chains with
m=3,4,5 monomer units are presented. Results correspond to 
two isotherms, namely $T^{*}=1$ and $T^{*}=2$. 

In fig.2  a comparison is made between simulation and theoretical
results for LJ chains with $m=3$. 
In fig.2a results are presented for the equation of state (EOS)
and in fig.2b for the internal energy. 
As  can be seen the theory describes quite accurately the
simulation results both for the EOS and for the internal energies.
When expanding the solid it was found that the solid melts at a certain pressure.
We observed this espontaneous melting for the two considered
temperatures ($T^{*}=1$ and $T^{*}=2$). The lines of fig.2 correspond
to Wertheim's TPT1 theory. Fig.2 illustrates that Wertheim's TPT1 works
quite well not only for the fluid phase, but also for the solid phase.
It is well known that Wertheim's TPT1 describes quite well the fluid
phase. Fig.2 show clear evidence that Wertheim's TPT1 describes
quite well also the solid phase.
In fig.3 a comparison is made between simulation and theoretical
results for LJ chains with $m=4$. 
In fig.3a results are presented for the equation of state (EOS)
and in fig.3b for the internal energy.
Again the theory performs very well for both properties.
For the two considered temperatures the solid phase melted into an isotropic
fluid when decreasing the pressure. 
Finally in fig.4 a comparison is made between simulation and theoretical
results for LJ chains in the solid phase with $m=5$. 
In fig.5a results are presented for the equation of state (EOS)
and in fig.5b for the internal energy. The theory one again is successful in describing 
the simulation results. 
For the two considered temperatures the solid phase melted into an isotropic
fluid when decreasing the pressure. 

 For each of  the chains considered in this work the agreement 
between theory and simulation for $T^{*}=1$ is as good as 
that for $T^{*}=2$ thus indicating  that the theory does  not  deteriorate
with respect to  variations in the  temperature (at least for the range of temperatures
considered here). 

 Good agreement between theory and simulation 
has previously been seen for the LJ dimer (m=2) in the solid phase\cite{vega2002}.
However, it is certainly gratifying to observe that the theory continues to perform well 
with increasing chain length.
 The message of figures 2-4 is that Wertheim's TPT1 theory
can be used with confidence to predict the thermodynamic
properties, both  for  the EOS and internal energies of LJ chains in the solid
phase. 

It would be  interesting to see if Wertheim's
theory is able also able to accurately  predict  the free energies
of LJ chains in the solid phase. The determination of 
free energies of solid from computer simulation requires
special techniques (Einstein crystal calculations\cite{ladd,frenkelsmit,polson2}).
At this point it is possible to  mention that for $m=2$ the 
comparison between free energies of the solid phase as 
obtained from Wertheim's TPT1 and from Einstein crystal calculations
is quite good \cite{unpublished}. Free energy calculations of LJ chains
would certainly present an interesting problem for future work. 

 One of the consequences of Wertheim theory is the existence
of asymptotic limits\cite{vegamacdowell2001,vega2002} of the 
thermodynamic properties when plotted
as a function of the number of monomer units per unit of volume
(instead of plotting
them as a function of the number of molecules per unit of volume). 
Let let us define the monomer number density $\rho_{m}$ 
as $\rho_{m}=\rho m$. 
 In Wertheim's formalism, the compressibility 
factor $Z$ and residual internal energy $U/(NkT)$ can be written 
as \cite{vegamacdowell2001} 

\begin{equation}
X(\rho_{m},T,m)=X_{1}(\rho_{m},T)+mX_{2}(\rho_{m},T),
\label{linear}
\end{equation}

\noindent where $X$ stands for 
any of the thermodynamic properties (Z or $U/(NkT)$ ).
For sufficiently large values of $m$, Eq.\ref{linear}
can be written as:

\begin{equation}
 p^{*}=p/(\epsilon/\sigma^{3})  = kT \rho^{*} m Z_{2}(\rho_{m},T) =
                           kT \rho_{m}^{*} Z_{2}(\rho_{m},T) 
\label{linearp}
\end{equation}

 and 
\begin{equation}
 U/(kT) =   N m  U_{2}( \rho_{m},T) =
    N_{m} U_{2}(\rho_{m},T) 
\end{equation}
\begin{equation}
 U/(N_{m}kT)  = U_{2}(\rho_{m},T) 
\label{linearu}
\end{equation}

 According to Eq.(\ref{linearp}) for very long
chains ($m$ sufficiently large) , the reduced pressure is a function 
of the reduced temperature and of the reduced monomer number density
only. In view of this  a plot of $p^{*}$ versus $\rho_{m}$ for a certain
temperature should yield a universal curve (regardless of the length of the
chain $m$).
 Also according to Eq.(\ref{linearu}) for very long
chains ($m$ sufficiently large) , the residual internal energy per monomer unit
depends only on the temperature and on the reduced monomer density.
Therefore a plot of $U/(N_{m}kT)$ versus $\rho_{m}$ for a certain
temperature should also yield a universal curve.
Obviously the chains considered in this work are rather short ($m=3,4,5$) and do not constitute
large values of $m$. 
However it is
worth checking whether this prediction of Wertheim's TPT1 holds for the
simulation results of this work.
In fig.5a the reduced pressure is plotted as a function of the reduced monomer
number density. In fig.5b the residual internal energy is plotted as a function 
of the reduced monomer number density. It can be seen that although the results
depend on the length of the chain the presence of an asymptotic limit is 
clearly visible. The results of fig.5 support the prediction of TPT1 on the
existence of an asymptotic limit and illustrate that even a chain of 
m=5 monomer units is not too far from such a limit. 

 This work has been devoted to fully flexible linear LJ chains and all the
results presented so far correspond to that model. 
We shall now present an different model that shall be denoted as 
the "linear rigid" chain. 
In the linear rigid chain 
the bond length, bond angles and internal degrees of freedom 
are fixed.  This linear rigid LJ chain is formed by $m$ LJ monomer units with bond length
$L=\sigma$ in a linear rigid configuration. 
Notice that the Hamiltonian of a system of
fully flexible LJ chains and that of the linear rigid LJ chain are different.
In fact in the first model there are intra-molecular interactions (between non
bonded monomer units) whereas these intra-molecular interactions are missing
in the "linear rigid chain" (or one can say more generally that the intra-molecular
energy is simply a constant). Also in the first model the molecule can adopt
any bond angle or torsional state reflecting the fact the molecule is indeed
flexible whereas these degrees of freedom are
frozen in the "linear rigid chain".  
Wertheim's TPT1 theory does not distinguish between fully flexible and "linear
rigid" chains and therefore it predicts the same EOS for
for a fully flexible chain and for a fully rigid chain.
This surprising results holds very well in the fluid phase, both for hard sphere
chains\cite{boublik1990} and for LJ chains. In fact in the fluid phase fully flexible and fully
rigid chains present a quite similar (although not identical) EOS. Is this  
also true in the solid phase?
It has been shown  recently \cite{lths,vegapre} that the similarity between fully flexible and fully
rigid linear chains does not hold in the solid phase. In fact in the case of 
hard sphere chains the  EOS of fully flexible chains differs considerably with that  of 
 "linear rigid" chains. 
Given that Wertheim's TPT1 describes very well
the EOS of fully flexible hard sphere chains in the solid phase and fails 
completely in describing the EOS of "linear rigid" chains in the solid phase, one may 
inquire as to the situation
for LJ chains in the solid phase. To illustrate whether flexible and 
rigid LJ chains present the same or different EOS in the solid phase we have
performed NpT simulation for linear rigid LJ chains. Details are similar to 
those previously presented for flexible chains. Two difference are encountered.
Firstly the initial configuration 
of the solid is not a random configuration 
but an ordered one. 
The ordered structure used is the one   denoted as
CP1 in Ref.\cite{vega1992a}. This CP1 structure is 
similar to that proposed
by Polson and Frenkel for LJ semi-flexible chains.
The second difference is that we 
used anisotropic NpT scaling in the way described
in Ref.\cite{rahman,rao} This 
is important since our simulation box (and the symmetry of the unit
cell) is no longer cubic.
 
In Table IV simulations results for "linear rigid" LJ
chains with $m=3$ are presented. 
In fig.6a the EOS for $m=3$ as obtained from Wertheim's TPT1, and from the
Monte Carlo results of this work for fully flexible and linear rigid chains
are presented. The conclusion is clear. Although at very high densities flexible
and linear rigid chains present a quite similar equation of state, differences
are clearly visible at smaller densities within the solid phase.
The linear rigid chain tends to hava a higher density for a certain pressure.
A second significant difference was that for the linear rigid chaini model
the solid phase was mechanically stable up to zero pressure. That provides
some evidence that the triple point temperature of linear rigid chains may appears at 
higher temperatures than this of flexible chains. Free energy calculations
will be necessary to check this point. In fig.6b the internal energy for 
$m=3$ is presented for both flexible and linear rigid chains. Again differences
between flexible and linear rigid chains are clearly visible. For a certain
density the internal energy of the linear rigid chain is lower than this of
the flexible chain.
To analyze in more detail if differences between flexible and linear rigid
chains increases with the length of the chain, we have performed 
simulations also for linear rigid chains with $m=5$. Results are presented
in Table V. In fig.7 a comparison is made between flexible and linear rigid
chains with $m=5$. In fig.7a for the EOS and in fig.7b for the internal
energy. Differences between flexible and linear rigid chains are quite
evident. Differences are larger than for the model with $m=3$ indicating
that the differences between flexible and linear rigid chain in the solid
phase, increases with the length of the chain. Only at very high densities
(when approaching the "close packing") differences are found to be small. 

In summary we have found  that for LJ chains in the solid phase, fully flexible
and fully rigid molecules present  markedly different equations of state.
This has already been observed for  hard sphere chains, and the work here serves
as a further  illustration that
the same occurs again for LJ chains. Wertheim's TPT1 describes very well
the EOS of fully flexible chains but fails to describe the behavior of
of linear rigid LJ chains. One may conclude that the similarity between
fully flexible and "linear rigid" chains is broken in the solid phase.
Notice that the "close packed" density in the solid phase is the same 
for fully flexible and for linear rigid chains, so that a different packing
density is not the reason for the differences found in fig.6 and in fig.7.

 Let us just finish by analizing an interesting issue. Although it is clear
that for flexible chains the equilibrium solid structure is a disordered
solid, and for linear rigid chains the equilibrium solid structure is an 
ordered one, one may ask how would be the properties of a flexible chain
if arranged in an ordered solid. Malanoski and Monson \cite{malanoskichains} considered this
aspect for flexible hard chains. Malanoski and Monson showed that
the stable solid phase for hard flexible chains is a disordered solid \cite{malanoskichains}.
They also analized the behaviour of a flexible hard chain
in an ordered solid. They found that for hard flexible chains at a certain density,
the ordered solid gave a higher pressure than the disordered one. 
To analyze if this result holds also for LJ chains we decided to perform 
NpT simulations (with anisotropic scaling) of flexible LJ in an ordered
CP1 solid structure. Results for this flexible LJ chain in an ordered solid
structure are presented in Table VI. In fig.7 the results of these simulations
are also presented (they are labeled as flexible chain ordered solid).
As it can be seen in fig.7a for a certain density the pressure of the flexible chain
in the ordered solid is slightly higher than this of the flexible chain 
in the disordered solid. Differences are not too large but clearly visible.
Our results agree with those of Malanoski and Monson in the sense that
they yield higher pressures for the ordered solid with respect to the disordered solid.
A simple sumary of Fig.7a would be to say that flexible and linear rigid chains
present a quite different EOS in the solid phase and that differences between
different solid structures of the flexible chains (disordered or ordered) are 
much smaller) at least for the case considered here ($m=5$). 
In fig.7b results are presented for the internal energy. It is seen that
for the flexible chain, the ordered solid present a lower internal energy than this
of the disordered solid. The reader should not take that as an evidence of
the greater stability of the ordered solid with respect to the disordered
solid for flexible chains. Notice that
the disordered solid has a significant stabilizing contribution to the free 
energy arising from the degeneracy entropy which is not present in the ordered
solid. The small lower value of the internal energy of the ordered solid can 
not compensate the lost of the large value of the degeneracy entropy (at least for
not to loo temperatures). 

\pagebreak

\section {CONCLUSIONS}
 In this work  computer simulations were performed for fully flexible LJ
chains in the solid phase. The solid structure considered is based
on a face centered cubic (fcc) close packing arrangement of atoms.
The initial configuration of the chains is obtained by randomly distributing
 the molecular bonds of the chains. 
The NpT simulations of this work provide values of the EOS and internal  
energy of the LJ chains in the solid phase. The chains considered
here were formed by $m=3,4,5$ monomer units. 
The performance of Wertheim's TPT1 theory as extended recently to the solid
phase was analyzed. It was found that Wertheim's TPT1 provides a very
accurate description of the equation of state and residual internal
energies of LJ chains. In previous work we found good agreement for
the simplest case $m=2$ and here it is shown that the agreement is
also quite good for longer chains. The performance of the theory 
does not seem to deteriorate with the length of the chain (at least
for the lengths considered here). The results of this work provide
further evidence that Wertheim's TPT1 can be used with confidence
for fully flexible LJ chains in the solid phase.
It was also found that for a fixed temperature the EOS and internal energies of LJ chains
tend to asymptotic limits when plotted as a function of 
the monomer number density. This behavior is suggested by Wertheim's TPT1
and our simulation results are fully consistent with that behavior. 
Finally some simulations were also performed for "linear rigid" LJ chains
in an ordered solid (CP1 structure). We found differences in the thermodynamic
properties of fully flexible and "linear rigid" chains in the solid phase. 
Therefore the similarity in thermodynamic properties between fully flexible 
and "linear rigid" chains predicted by Wertheim's TPT1 does not hold in 
the solid phase (although it holds quite well in the isotropic phase).
This work shows that Wertheim's TPT1 can be used to accurately describe  the
thermodynamic properties of fully flexible chains, both in the fluid and in 
the solid phase.

\section*{Acknowledgements}
Financial support is due to project number BFM-2001-1420-C02-01
of  Spanish
DGICYT (Direcci\'on General de Investigaci\'on Cient\'{\i}fica
y T\'ecnica).

\pagebreak

\pagebreak

FIGURE CAPTIONS

Fig. 1. The EOS of fully flexible hard sphere chains with $m=4$ in 
the solid phase as 
obtained from computer simulations of this work (diamonds), 
from computer simulations of Malanoski and Monson (full line)
and from Wertheim's TPT1 (dashed line) as proposed and implemented 
in Ref.\cite{vegamacdowell2001}. 
The solid structure is that
of an fcc arrangement of atoms with a random assignment of
bonds. 

Fig. 2. Properties of fully flexible LJ chains with $m=3$ 
as 
obtained from computer simulations of this work (symbols)
and from Wertheim's TPT1 
for the fluid\cite{johnson2} and for the solid\cite{vega2002} 
phases (lines). Results were obtained
for two reduced temperatures, $T^{*}=1$ and $T^{*}=2$.
The solid structure is that
of an fcc arrangement of atoms with a random assignment of
bonds. 
a) Results for the reduced pressure, $p^{*}=p/(\epsilon/\sigma^{3})$
as a function of the reduced number density of chains. 
b) Results for the residual internal energy  $U/(N\varepsilon)$
as a function of the reduced number density of chains. 

Fig.3. As in fig.2 for LJ chains with $m=4$.
a) Results for the reduced pressure. 
b) Results for the residual internal energy

Fig.4. As in fig.2 for LJ chains with $m=5$.
a) Results for the reduced pressure. 
b) Results for the residual internal energy.

Fig.5  a) Reduced pressure $p^{*}=p/(\epsilon/\sigma^{3})$ as a function 
of the monomer density for fully flexible LJ chains.
Symbols, simulation results of this work. Lines, Wertheim's TPT1. 
b) Residual internal energy $U_{m}/(N\varepsilon)$ as a function 
of the monomer density for fully flexible LJ chains.
Symbols, simulation results of this work. Lines, Wertheim's TPT1. 

Fig.6 Properties of disordered solid formed by m=3 LJ flexible chains
(triangles down) and of 
ordered solid formed by m=3 LJ rigid chains (triangles up).
Theoretical prediction from Wertheim's TPT1 (solid line).
a) Results for the reduced pressure.
b) Results for the residual internal energy. 

Fig.7 Properties of disordered solid formed by m=5 LJ flexible chains
(triangles down), of ordered solid formed by LJ m=5 rigid chains (squares)
and of ordered solid formed by LJ m=5 flexible chains(circles).
Theoretical predictions from Wertheim's TPT1  (solid line).
a) Results for the reduced pressure.
b) Results for the residual internal energy. 

\pagebreak


\begin{table}
\caption
{ Simulation results for fully flexible LJ chains in the 
solid phase with m=3. }
\label{mcm3}
\begin{tabular}{cccc}
\hline
\hline
    $T$  & $p^{*}$  & $\rho^{*}$     & $U/(N\varepsilon)$    \\
\hline
\\
1 &            60.0 &          0.4667 &         -11.584 \\ 
1 &            50.0 &          0.4571 &         -13.638 \\ 
1 &            40.0 &          0.4455 &         -15.686 \\ 
1 &            30.0 &          0.4312 &         -17.642 \\ 
1 &            20.0 &          0.4127 &         -19.379 \\
1 &            10.0 &          0.3867 &         -20.571 \\ 
1 &             8.0 &          0.3795 &         -20.640 \\ 
1 &             6.0 &          0.3710 &         -20.636 \\ 
1 &             4.0 &          0.3597 &         -20.427 \\ 
1 &             2.0 &          0.3425 &         -19.821 \\
\\ 
1 &             1.0 &          0.2979 &         -16.881 \\ 
1 &             0.6 &          0.2921 &         -16.579 \\ 
1 &             0.2 &          0.2855 &         -16.223 \\ 
\hline
2 &            60.0 &          0.4584 &         -10.151 \\ 
2 &            50.0 &          0.4474 &         -12.191 \\ 
2 &            40.0 &          0.4343 &         -14.163 \\ 
2 &            30.0 &          0.4176 &         -15.935 \\ 
2 &            20.0 &          0.3951 &         -17.287 \\ 
\\
2 &            10.0 &          0.3252 &         -15.238 \\ 
2 &             6.0 &          0.3004 &         -14.782 \\ 
2 &             0.2 &          0.1880 &          -9.511 \\ 
\\
\hline
\hline
\end{tabular}
\end{table}    
\pagebreak 
\begin{table}
\caption
{ Simulation results for LJ chains with m=4. }
\label{mcm4}
\begin{tabular}{cccc}
\hline
\hline
       $T$  & $p^{*}$  & $\rho^{*}$     & $U/(N\varepsilon)$    \\
\hline
\\
1 &            65.0 &          0.3534 &         -14.178 \\ 
1 &            60.0 &          0.3501 &         -15.532 \\ 
1 &            50.0 &          0.3432 &         -18.184 \\ 
1 &            40.0 &          0.3350 &         -20.821 \\ 
1 &            30.0 &          0.3240 &         -23.439 \\ 
1 &            20.0 &          0.3105 &         -25.818 \\ 
1 &            10.0 &          0.2915 &         -27.348 \\ 
1 &             8.0 &          0.2863 &         -27.432 \\ 
1 &             6.0 &          0.2797 &         -27.402 \\ 
1 &             4.0 &          0.2722 &         -27.222 \\ 
1 &             2.0 &          0.2609 &         -26.591 \\ 
1 &             1.0 &          0.2497 &         -25.582 \\ 
1 &             0.8 &          0.2459 &         -25.151 \\
\\
1 &             0.7 &          0.2259 &         -22.626 \\ 
1 &             0.4 &          0.2201 &         -21.965 \\ 
1 &             0.2 &          0.2178 &         -21.774 \\ 
\hline
2 &            65.0 &          0.3508 &         -11.409 \\ 
2 &            60.0 &          0.3443 &         -13.767 \\ 
2 &            50.0 &          0.3364 &         -16.376 \\ 
2 &            40.0 &          0.3268 &         -18.936 \\ 
2 &            30.0 &          0.3145 &         -21.287 \\ 
2 &            20.0 &          0.2974 &         -22.979 \\ 
2 &            10.0 &          0.2616 &         -22.462 \\ 
\\
2 &             8.0 &          0.2376 &         -20.127 \\ 
2 &             4.0 &          0.2150 &         -19.009 \\ 
2 &             1.0 &          0.1798 &         -16.129 \\ 
2 &             0.2 &          0.1531 &         -13.627 \\ 
\\
\hline
\hline
\end{tabular}
\end{table}    

\pagebreak
\begin{table}
\caption
{ Simulation results for LJ chains with m=5. }
\label{mcm5}
\begin{tabular}{cccc}
\hline
\hline
   $T$  & $p^{*}$  & $\rho^{*}$     & $U/(N\varepsilon)$    \\
\hline
\\
1 &            60.0 &          0.2806 &         -19.571 \\ 
1 &            50.0 &          0.2751 &         -22.840 \\ 
1 &            40.0 &          0.2688 &         -26.007 \\ 
1 &            30.0 &          0.2605 &         -29.074 \\ 
1 &            20.0 &          0.2495 &         -32.020 \\ 
1 &            10.0 &          0.2342 &         -34.033 \\ 
1 &             8.0 &          0.2297 &         -34.258 \\ 
1 &             6.0 &          0.2247 &         -34.262 \\ 
1 &             4.0 &          0.2189 &         -34.112 \\ 
1 &             2.0 &          0.2098 &         -33.269 \\ 
1 &             1.0 &          0.2020 &         -32.151 \\ 
1 &             0.8 &          0.1987 &         -31.680 \\ 
1 &             0.6 &          0.1881 &         -29.587 \\ 
\\
1 &             0.4 &          0.1776 &         -27.543 \\ 
1 &             0.2 &          0.1762 &         -27.354 \\ 
\hline
2 &            60.0 &          0.2760 &         -17.281 \\ 
2 &            50.0 &          0.2694 &         -20.552 \\ 
2 &            40.0 &          0.2620 &         -23.539 \\ 
2 &            30.0 &          0.2520 &         -26.545 \\ 
2 &            20.0 &          0.2393 &         -28.949 \\ 
2 &            10.0 &          0.2160 &         -29.202 \\
\\ 
2 &             8.0 &          0.1913 &         -25.250 \\ 
2 &             2.0 &          0.1586 &         -22.056 \\ 
2 &             0.2 &          0.1280 &         -17.691 \\ 
\\
\hline
\hline
\end{tabular}
\end{table}    

\pagebreak

\begin{table}
\caption
{ Simulation results for linear rigid  LJ chains in the 
solid phase with m=3 }
\label{mcrigidm3}
\begin{tabular}{cccc}
\hline
\hline
  $T$  & $p^{*}$  & $\rho^{*}$     & $U/(N\varepsilon)$    \\
\hline
1 & 60.0 & 0.4683  & -12.072\\
1 & 40.0 & 0.4459  & -16.650\\
1 & 30.0 & 0.4312  & -18.799\\
1 & 20.0 & 0.4131  & -20.695\\
1 & 10.0 & 0.3895  & -21.998\\
1 & 8.0 & 0.3834  & -22.134\\
1 & 6.0 & 0.3764  & -22.196\\
1 & 5.0 & 0.3723  & -22.190\\
1 & 4.0 & 0.3678  & -22.154\\
1 & 3.0 & 0.3628  & -22.080\\
1 & 2.0 & 0.3572  & -21.950\\
1 & 1.0 & 0.3507  & -21.753\\
1 & 0.6 & 0.3479  & -21.648\\
1 & 0.1 & 0.3441  & -21.498\\
\\
\hline
\hline
\end{tabular}
\end{table}    

\pagebreak

\begin{table}
\caption
{ Simulation results for linear rigid  LJ chains in the 
solid phase with m=5 }
\label{mcrigidm3}
\begin{tabular}{cccc}
\hline
\hline
  $T$  & $p^{*}$  & $\rho^{*}$     & $U/(N\varepsilon)$    \\
\hline
1 &            60.0 &          0.2824 &         -20.672 \\ 
1 &            40.0 &          0.2699 &         -28.128 \\ 
1 &            30.0 &          0.2615 &         -31.729 \\ 
1 &            20.0 &          0.2512 &         -34.979 \\ 
1 &            10.0 &          0.2384 &         -37.335 \\ 
1 &             8.0 &          0.2354 &         -37.624 \\ 
1 &             6.0 &          0.2319 &         -37.827 \\ 
1 &             5.0 &          0.2299 &         -37.893 \\ 
1 &             4.0 &          0.2277 &         -37.917 \\ 
1 &             3.0 &          0.2252 &         -37.903 \\ 
1 &             2.0 &          0.2224 &         -37.835 \\ 
1 &             1.0 &          0.2194 &         -37.708 \\ 
1 &             0.8 &          0.2187 &         -37.671 \\ 
1 &             0.1 &          0.2163 &         -37.521 \\ 
\\
\hline
\hline
\end{tabular}
\end{table}    

\pagebreak
\begin{table}
\caption
{ Simulation results for an ordered solid formed by flexible m=5 LJ chains.}
\label{mcrigidm3}
\begin{tabular}{cccc}
\hline
\hline
  $T$  & $p^{*}$  & $\rho^{*}$     & $U/(N\varepsilon)$    \\
\hline
1 &            60.0 &          0.2805 &         -20.077 \\ 
1 &            50.0 &          0.2743 &         -23.866 \\ 
1 &            40.0 &          0.2668 &         -27.457 \\ 
1 &            30.0 &          0.2579 &         -30.880 \\ 
1 &            20.0 &          0.2469 &         -33.762 \\ 
1 &            15.0 &          0.2400 &         -34.768 \\ 
1 &            10.0 &          0.2315 &         -35.427 \\ 
1 &             8.0 &          0.2269 &         -35.414 \\ 
1 &             6.0 &          0.2221 &         -35.409 \\ 
1 &             4.0 &          0.2165 &         -35.218 \\ 
1 &             2.0 &          0.2079 &         -34.272 \\ 
1 &             1.0 &          0.2025 &         -33.523 \\ 
1 &             0.2 &          0.1975 &         -32.778 \\ 
\\
\hline
\hline
\end{tabular}
\end{table}    

\pagebreak
\begin{figure}
\includegraphics[clip]{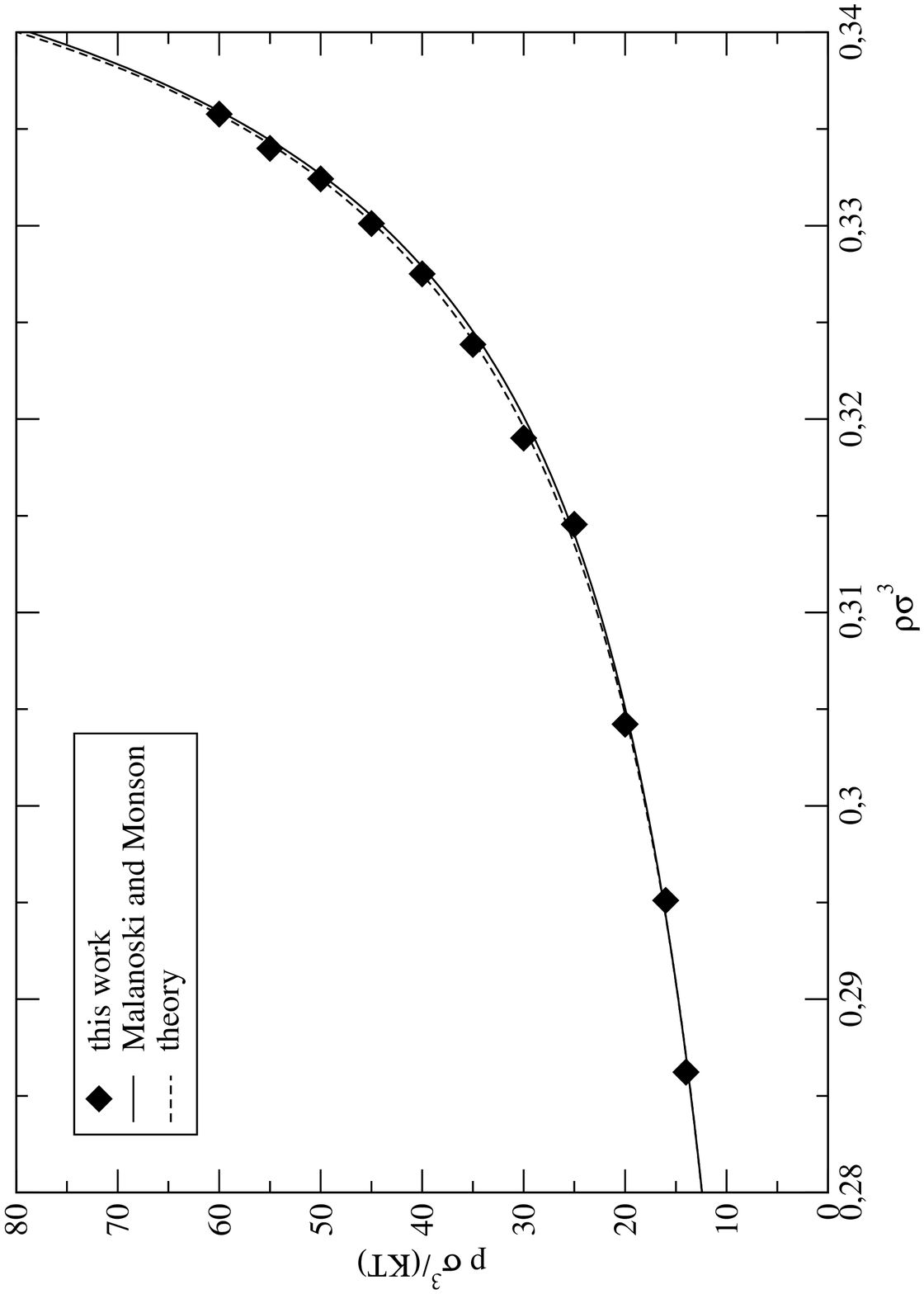}
\end{figure}

\pagebreak
\begin{figure}
\includegraphics[clip]{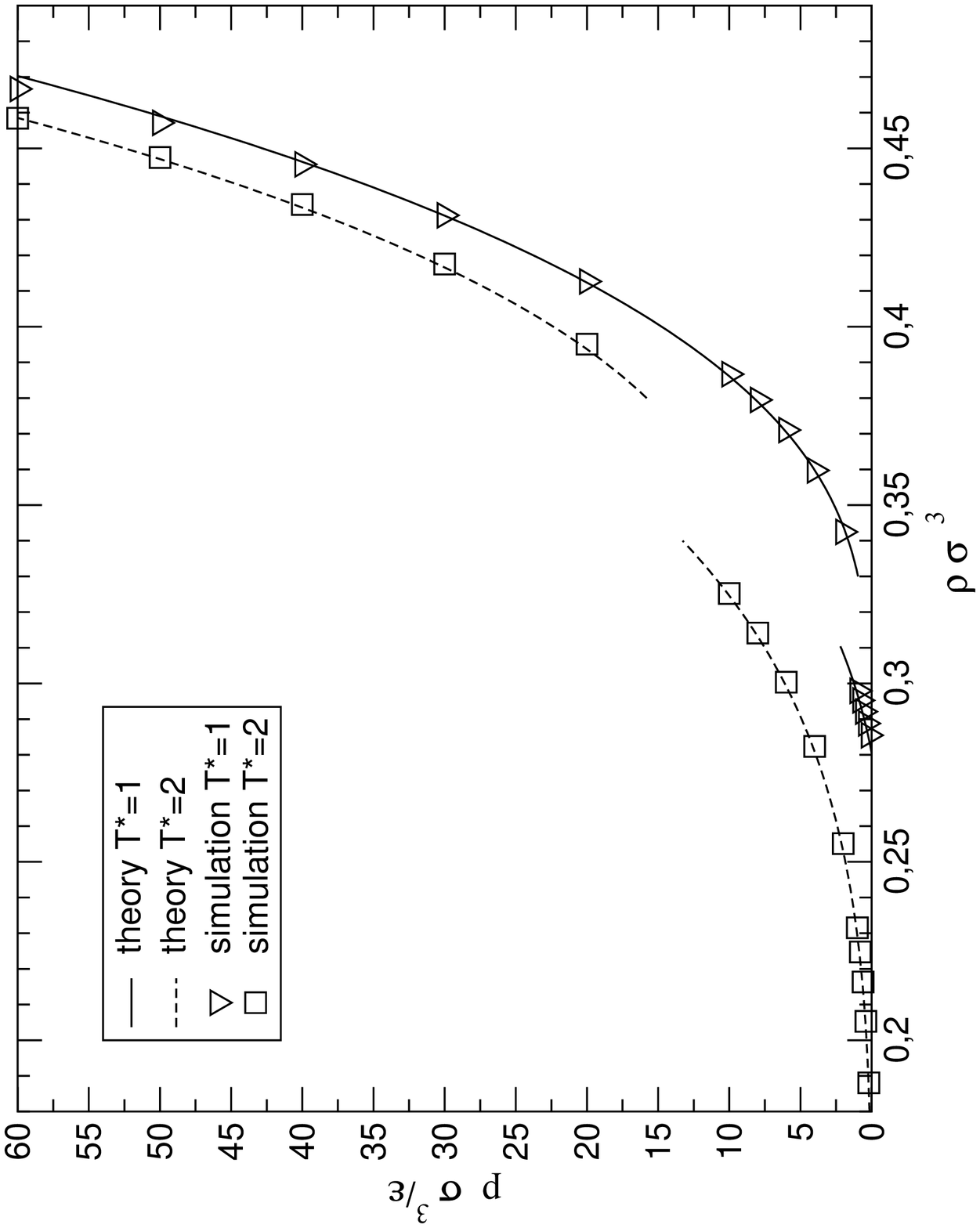}
\end{figure}
\pagebreak
\begin{figure}
\includegraphics[clip]{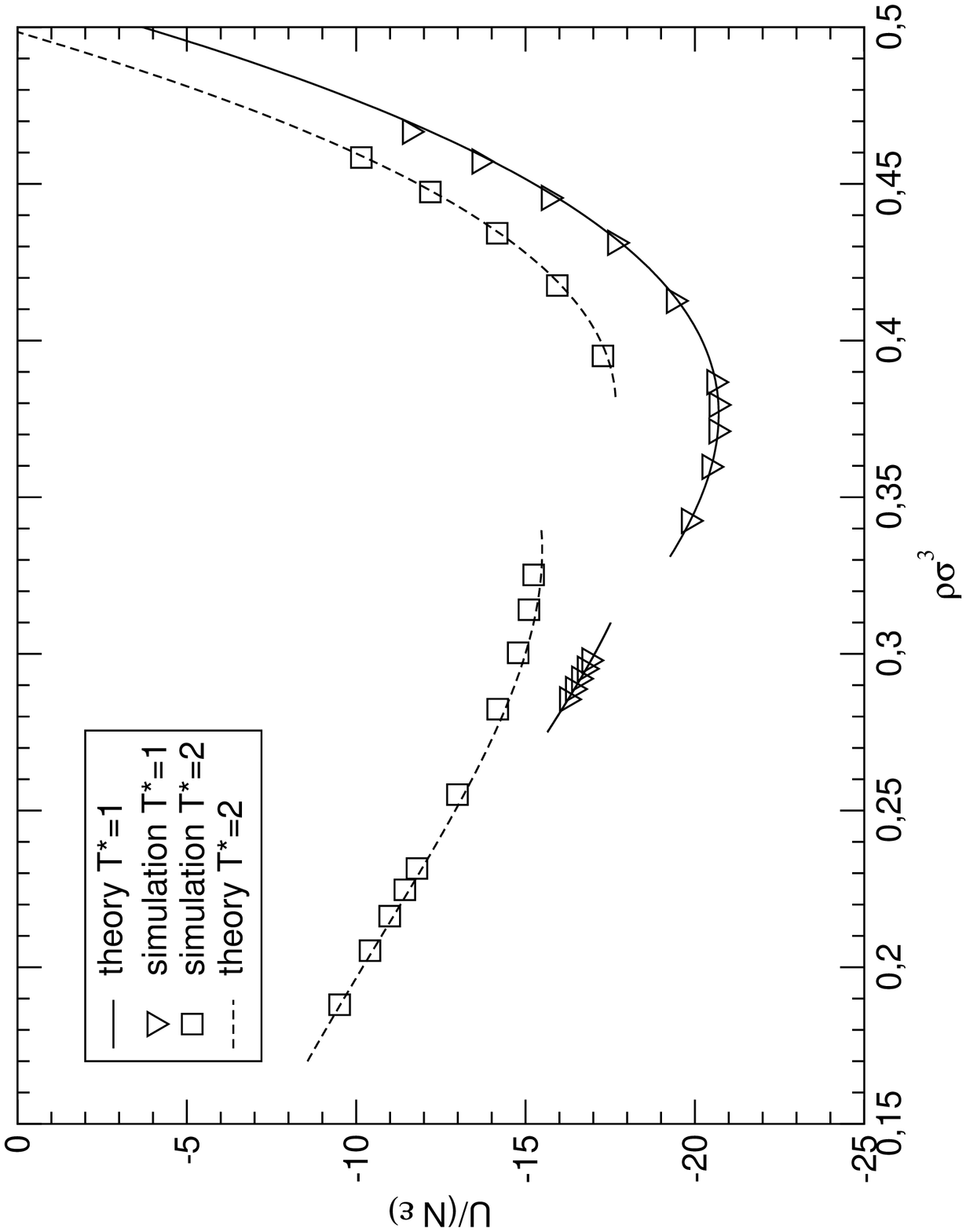}
\end{figure}
\pagebreak
\begin{figure}
\includegraphics[clip]{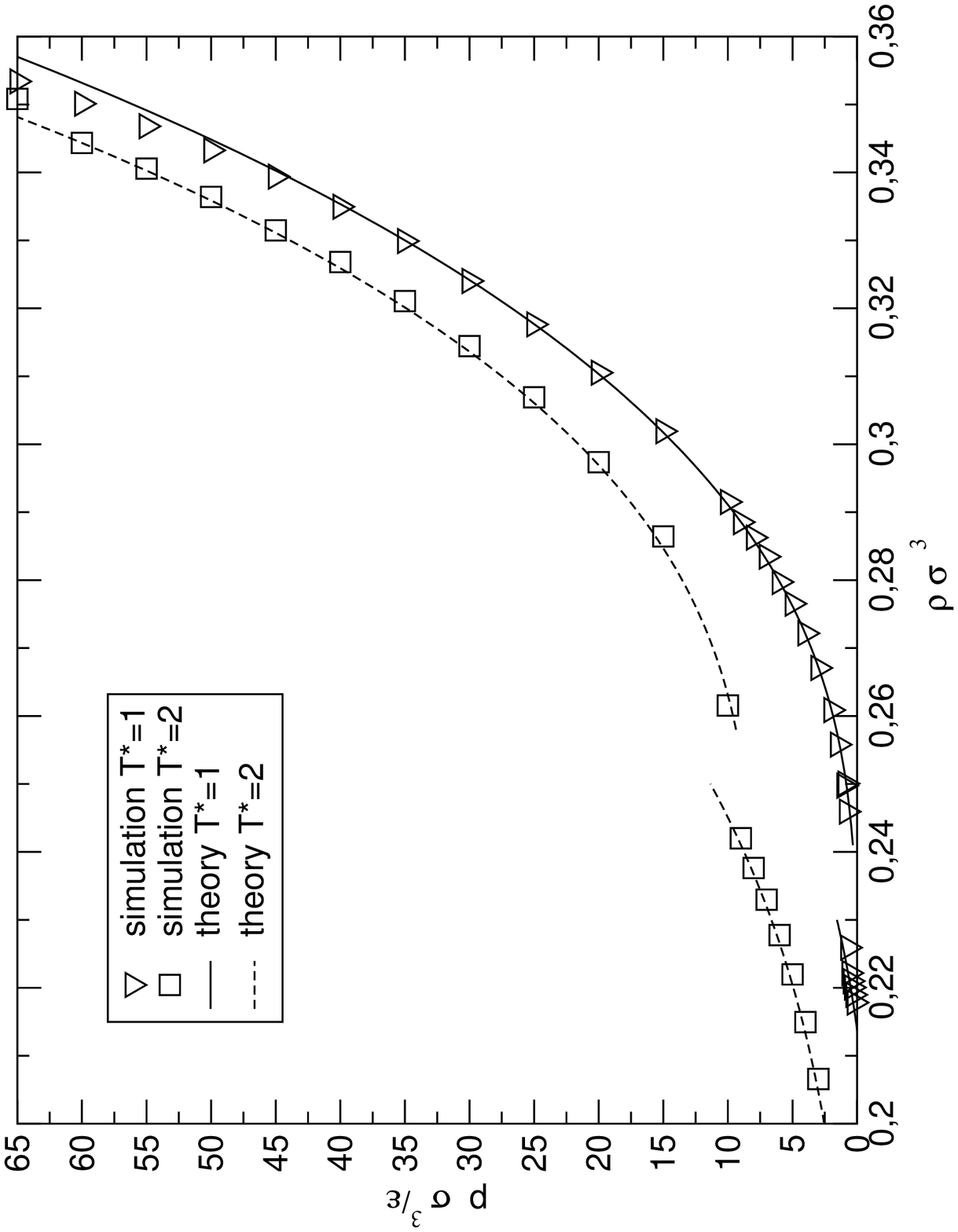}
\end{figure}
\pagebreak
\begin{figure}
\includegraphics[clip]{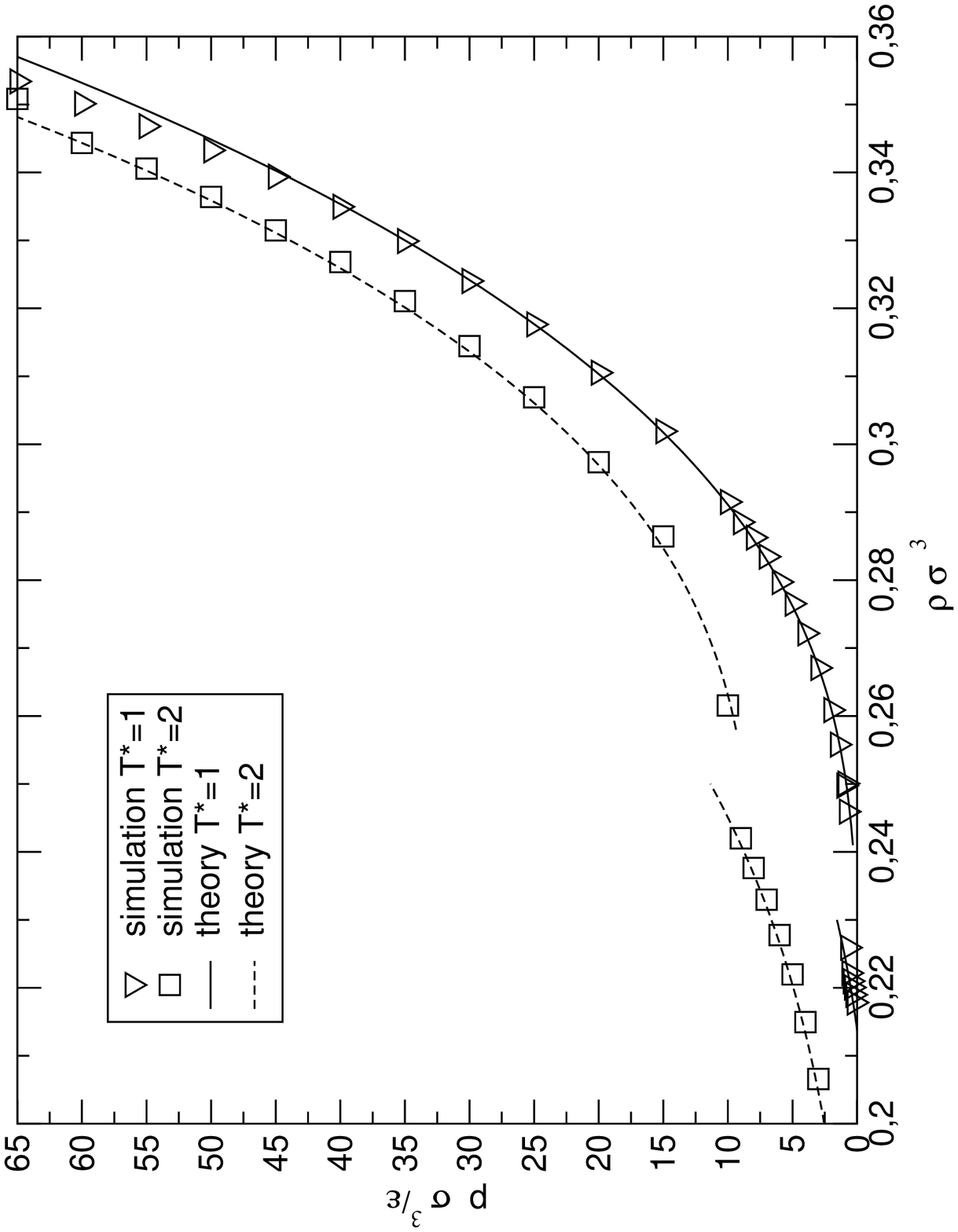}
\end{figure}
\pagebreak
\begin{figure}
\includegraphics[clip]{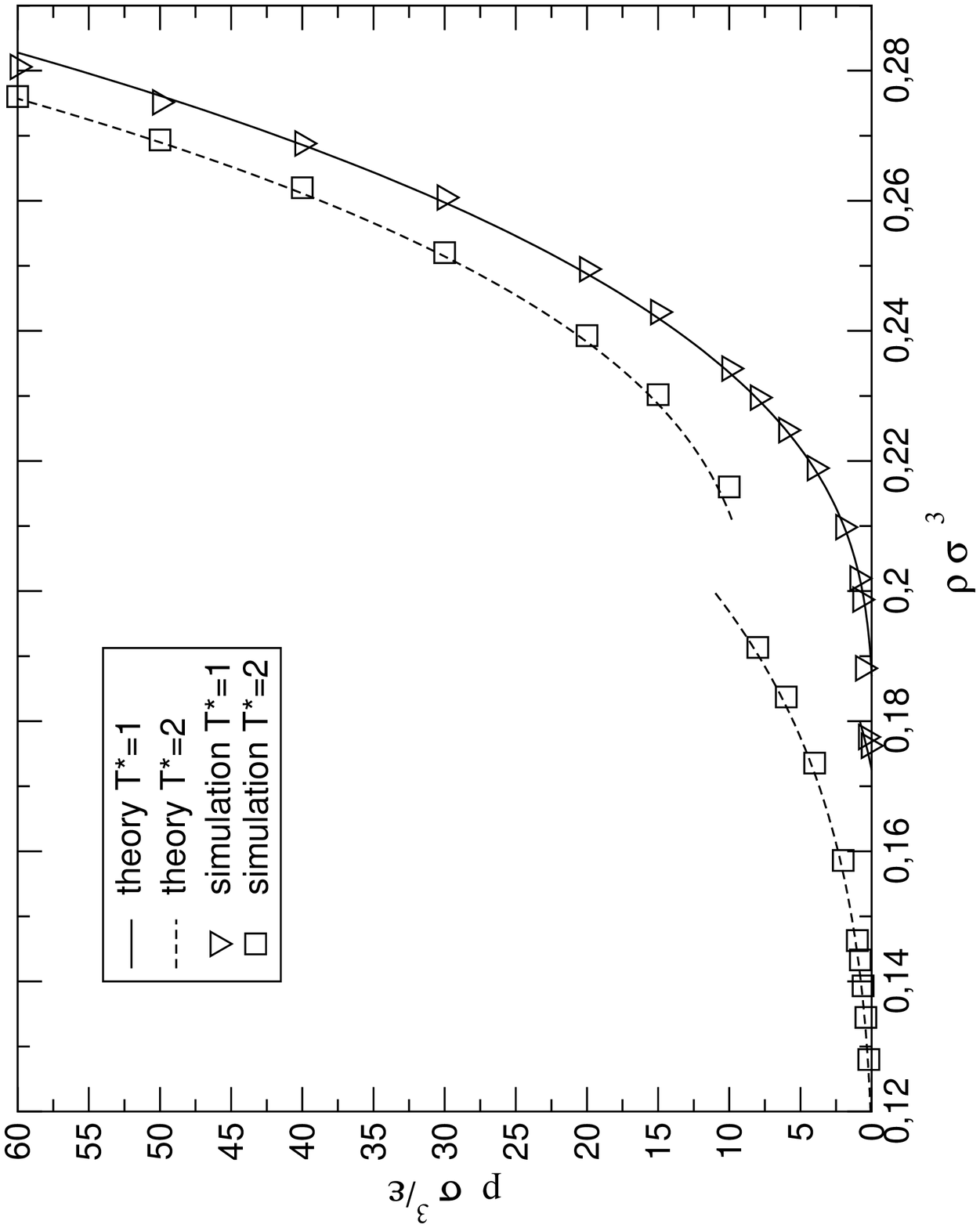}
\end{figure}
\pagebreak
\begin{figure}
\includegraphics[clip]{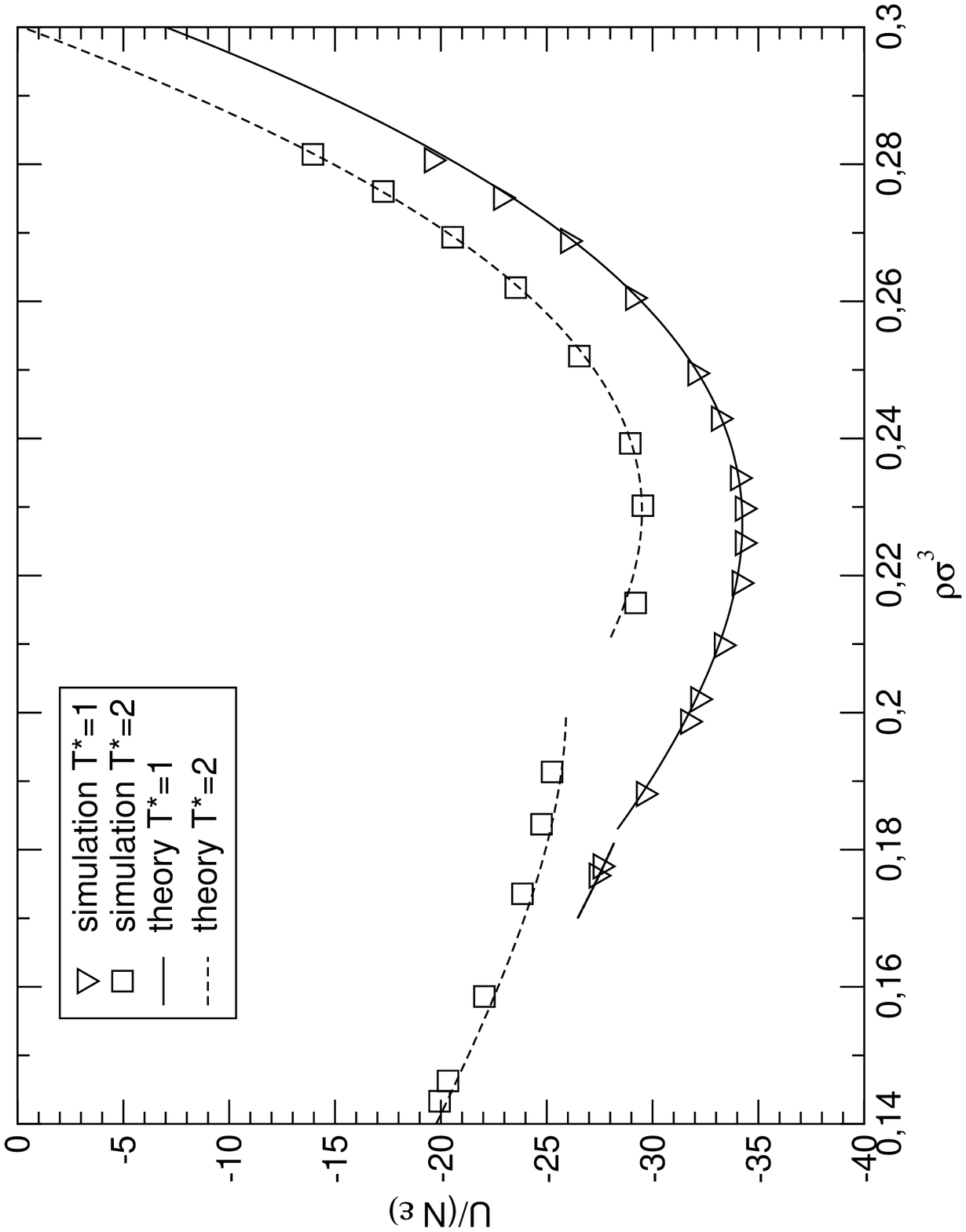}
\end{figure}
\pagebreak
\begin{figure}
\includegraphics[clip]{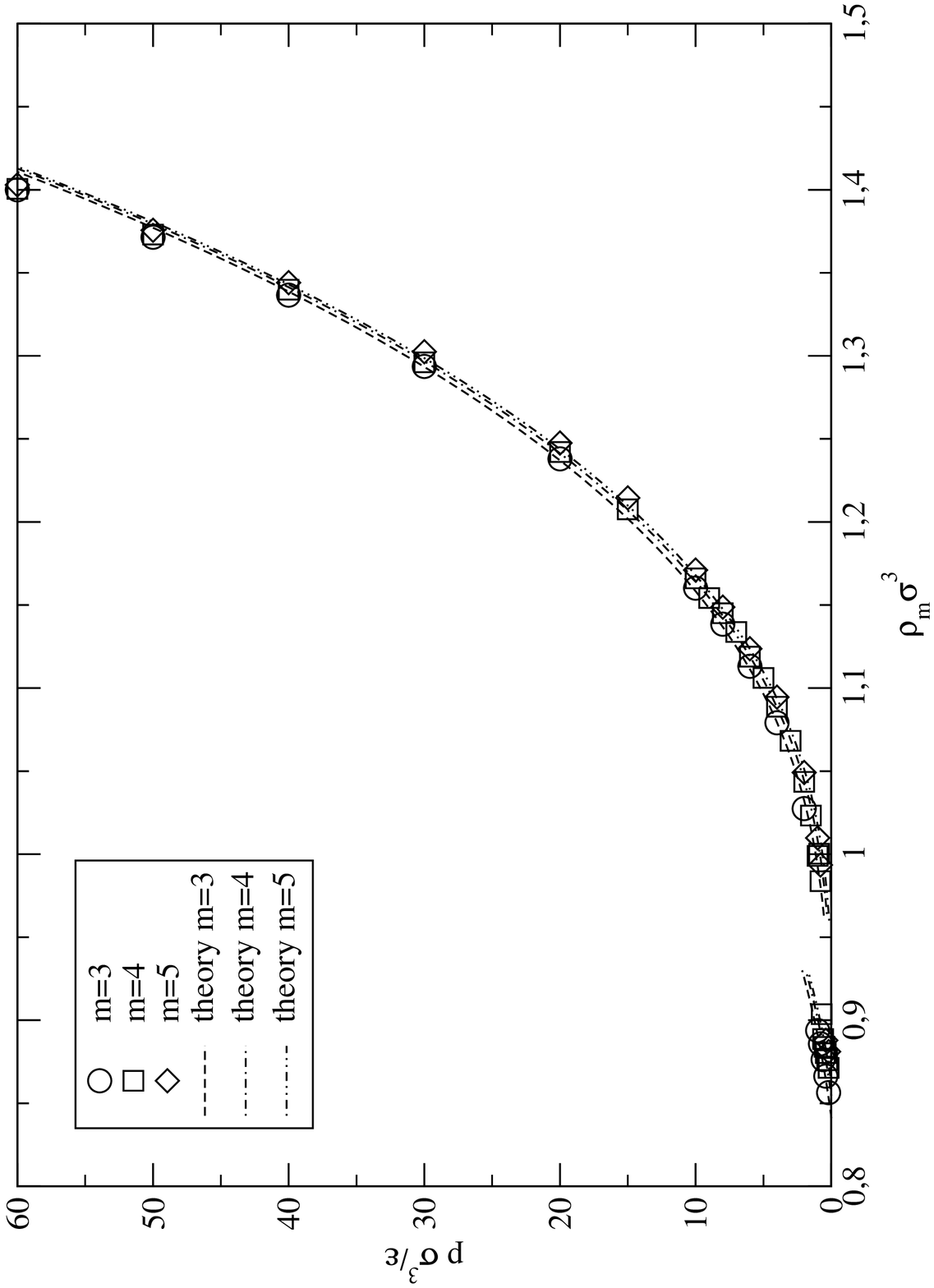}
\end{figure}
\pagebreak
\begin{figure}
\includegraphics[clip]{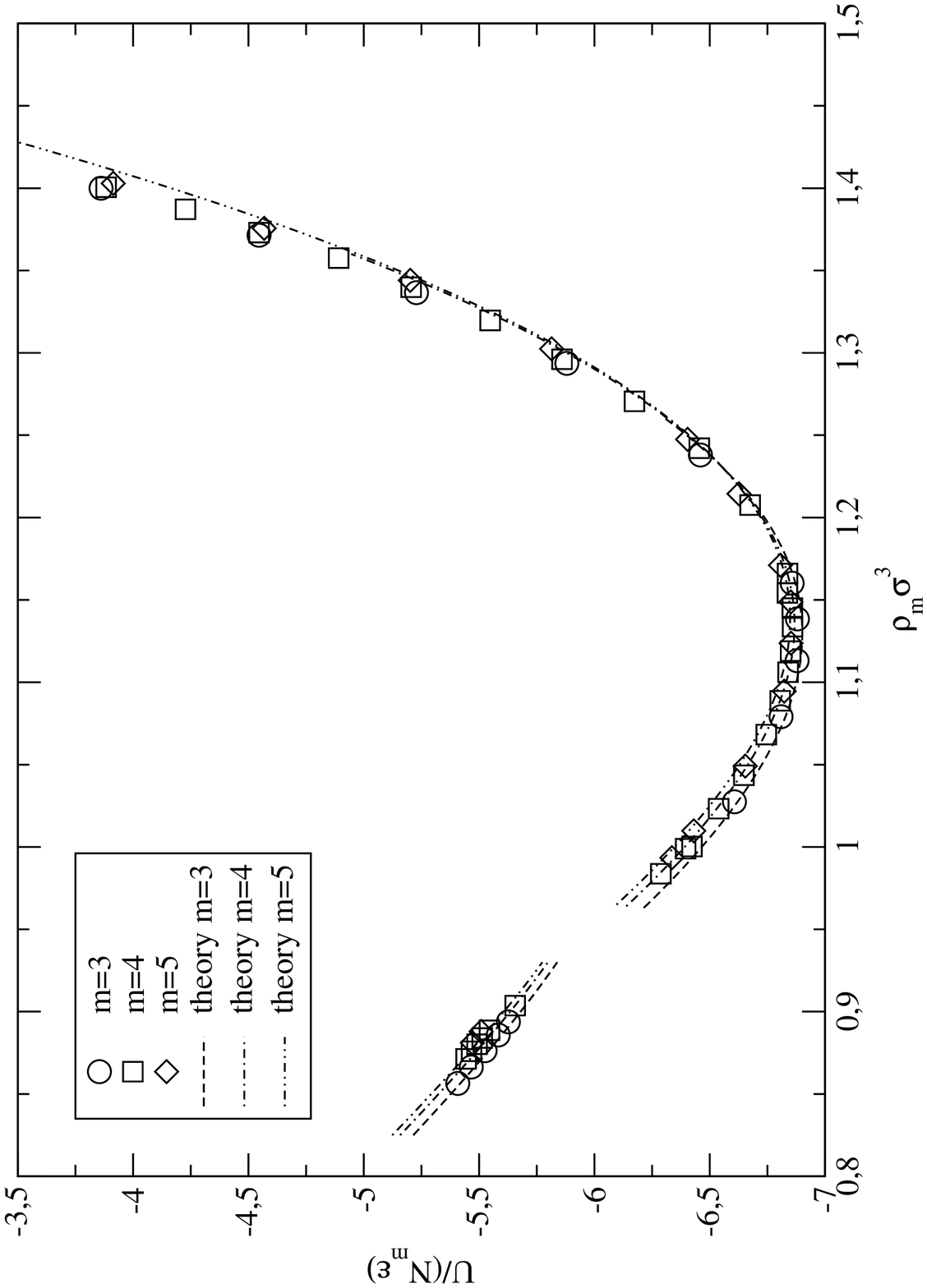}
\end{figure}
\pagebreak
\begin{figure}
\includegraphics[clip]{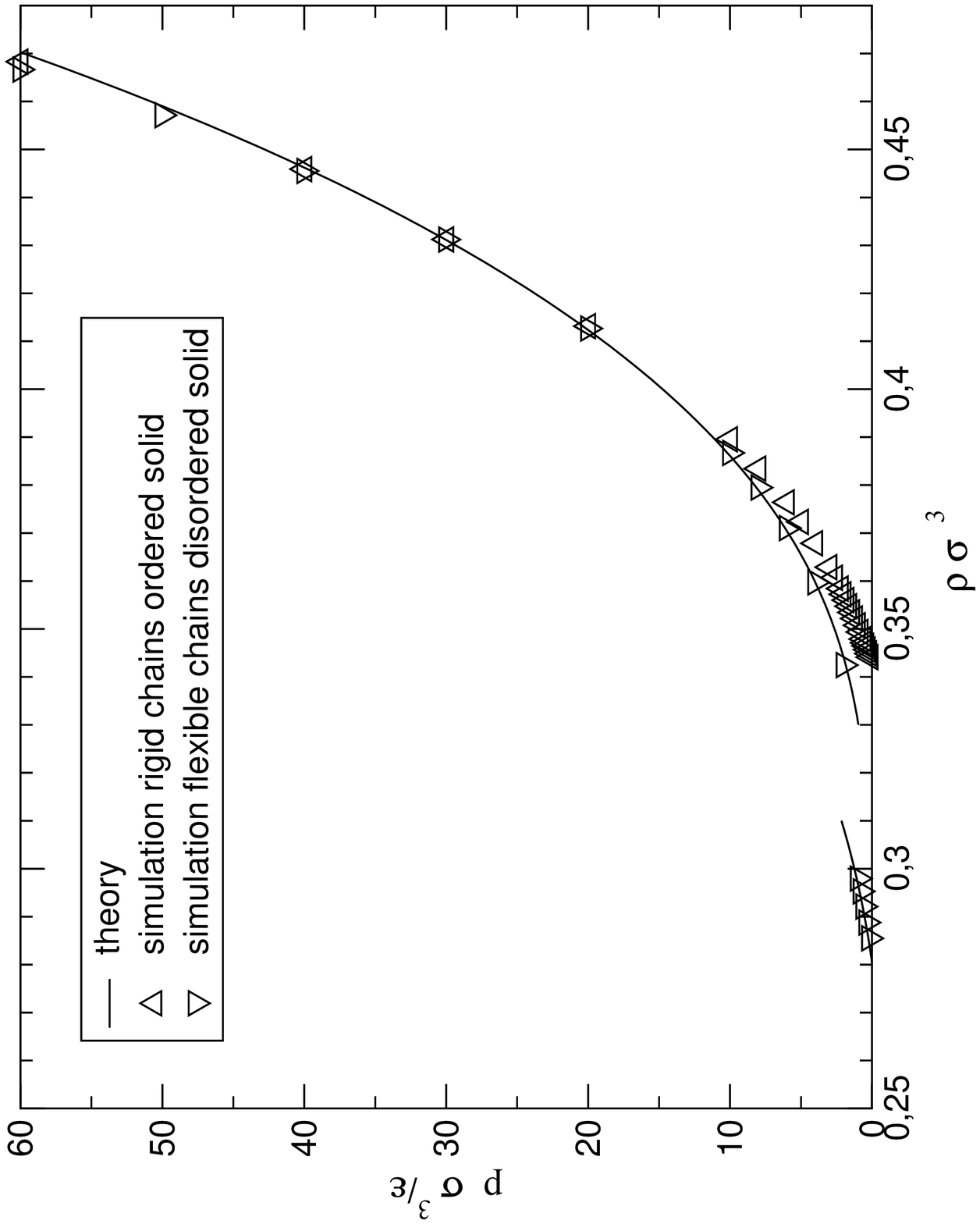}
\end{figure}
\pagebreak
\begin{figure}
\includegraphics[clip]{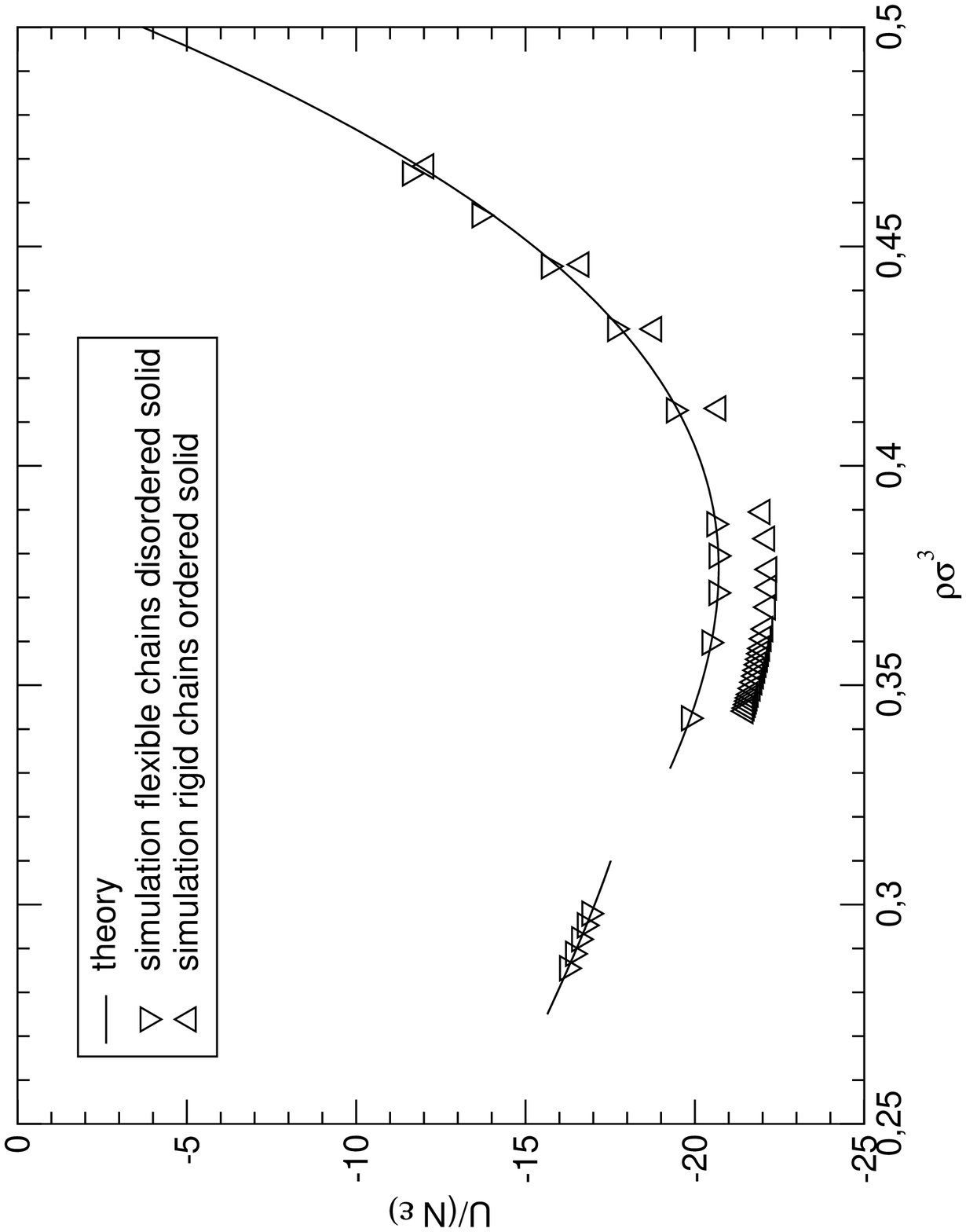}
\end{figure}
\pagebreak
\begin{figure}
\includegraphics[clip]{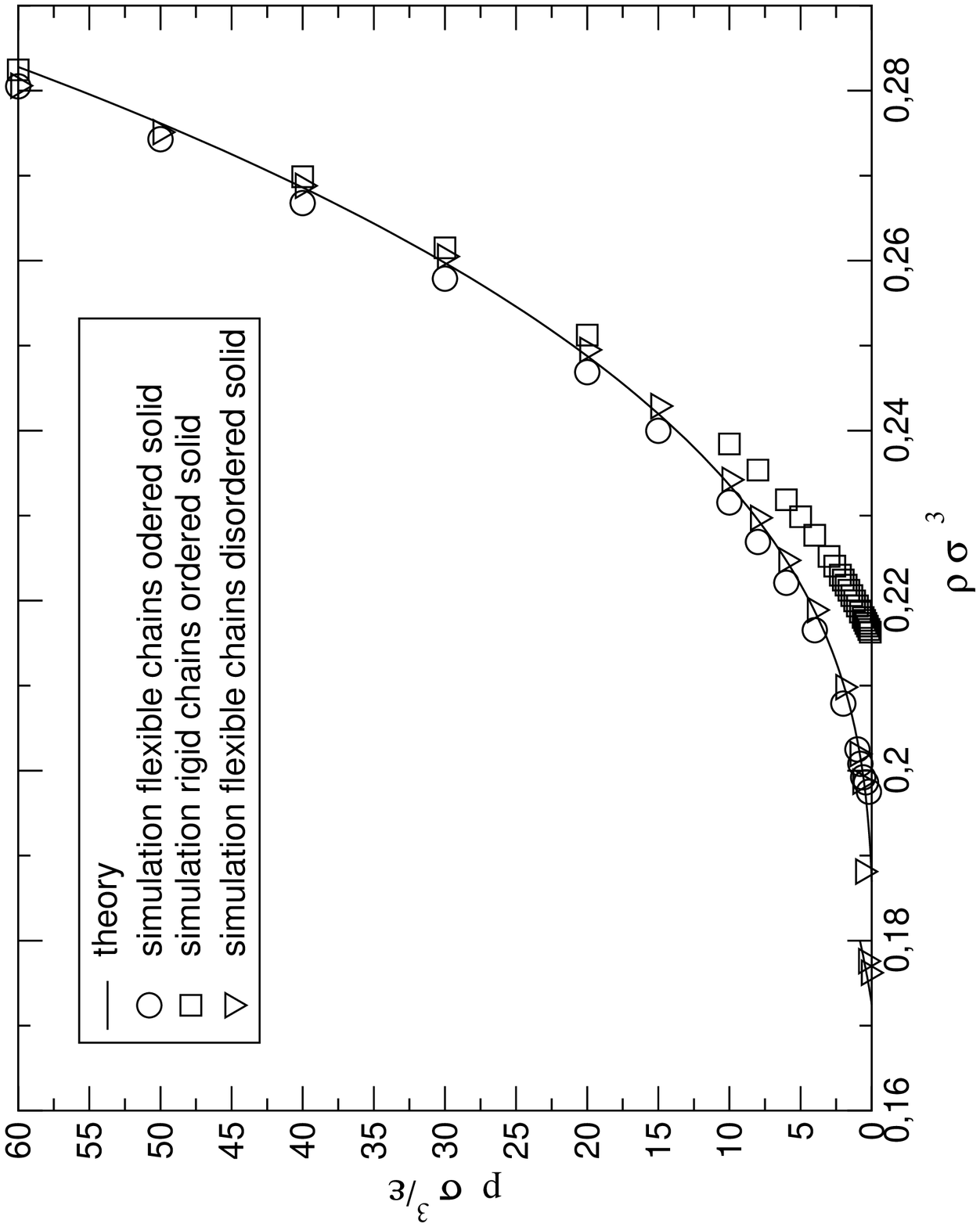}
\end{figure}
\pagebreak
\begin{figure}
\includegraphics[clip]{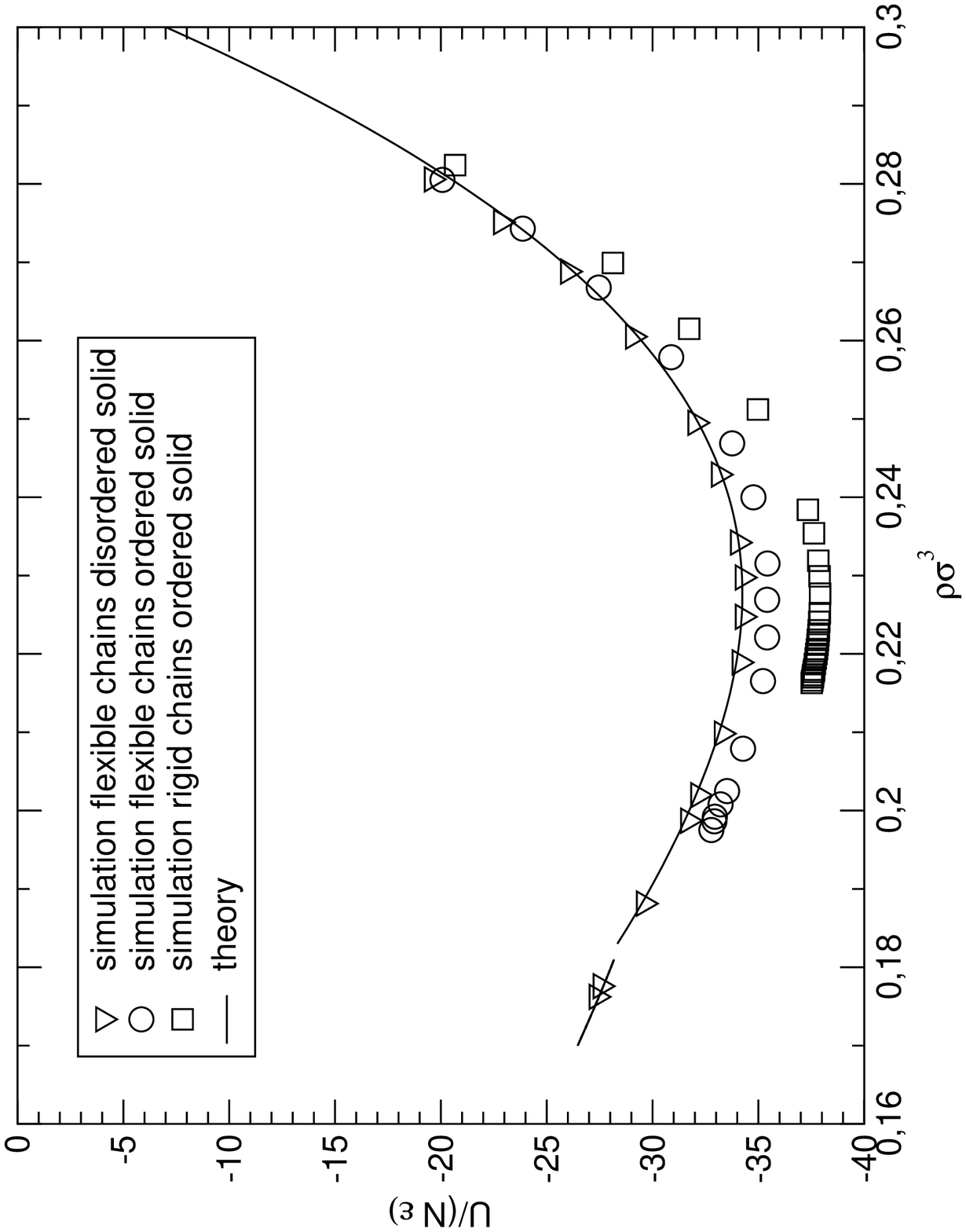}
\end{figure}

\end{document}